\documentclass[preprint,showpacs,preprintnumbers,amsmath,amssymb]{revtex4}

\usepackage{graphicx}
\usepackage{dcolumn}
\usepackage{bm}
\usepackage[dvips]{color}

\begin{document}

\title{Comments on "Quantum Monte Carlo calculations of the potential energy curve of the helium dimer"}

\author{Xuebin Wu$^{1}$}
\email{ganymede@hsit.edu.cn}
\author{Chenlei Du$^{1}$}
\author{Jianbo Deng$^{1}$}
\email{dengjb@lzu.edu.cn}
%
 \affiliation{ $^1$Institute of Theoretical Physics, Lanzhou University, Lanzhou 730000, People's Republic of China\\
}%

\date{\today}

\begin{abstract}
In a very recent paper[J. Chem. Phys. 128, 114308 (2008)], Springall
and co-workers reported Quantum Monte Carlo calculations on the
potential energy curve of the helium dimer. They argued that their
FN-DMC results are mostly within 1\% of the Hurly-Mehl curve[J. Res.
Natl. Inst. Stand. Technol. 112, 75 (2007)]. In this comment we show
that their results are wrong. Their reference-results computed from
Hurly and Mehl's curve are completely inaccurate, and also their
FN-DMC results are wrong and unauthentic.
\end{abstract}



\maketitle

In a very recent paper\cite{Springall}, Springall and co-workers
reported Quantum Monte Carlo calculations on the potential energy
curve of the helium dimer. They employed Slater-Jastrow form trial
wave functions and used the fixed node approximation for the fermion
nodal surface. In their work, they present the results of
variational Monte Carlo VMC and fixed node diffusion Monte Carlo
Carlo (FN-DMC) calculations for the helium dimer with atomic
separations in the range R=[0.9,7.4] a.u. Their FN-DMC result for
helium-helium interaction energy at R=5.6 a.u is very accurate,
predicted -10.89$\pm$0.17K or -10.96$\pm$0.15K, which are in very
good agreement with the current accepted values of around 11.00K.
But, it seems very strange that their other results at different
separated distances are significantly different from the well
published results. For separated distance at R=4.5 a.u, the well
published and accepted values are all around 58.40K(Selected results
are listed in TABLE
I)\cite{GQMC,CC,EQMC1,prl,ACPF,CCFCI,SAPT,sapt2,CCSDTQ,ccr12,ECG,ccsdt},
 but their FN-DMC result is 49.76$\pm$0.29K, there is about 9K error;
for the separated distance at R=4.3 a.u, the corresponding well
published and accepted results are all around
118.0K\cite{ECCC,1,2,3,4,5}, but their FN-DMC result is
102.24$\pm$0.32K, this is nearly 16K error compare to the well
published and accepted results. These errors are very large in such
accurate Quantum Monte Carlo calculations, the errors are compare
to, even large than ground date interaction energy around 11.00K at
R=5.6 a.u. It is impossible and unbelievable that a benchmark
method(Quantum Monte Carlo) could give a accurate interaction energy
at one distance(R=5.6 a.u) but give inaccurate results(so large
errors) at other neighbor distances.

To investigate the possible error in their calculations, we
carefully performed systemic Quantum Monte Carlo calculations for
helium dimer\cite{our}. As is shown in the TABLE II, our results are
in excellent agreement with other well published and accepted
results. Even more surprising, in their origin paper, their listed
results computed from Hurly and Mehl's \cite{Hurly} analytic
potential for the helium dimer are also significantly different from
the well published results. We then carefully re-computed the Hurly
and Mehl's \cite{Hurly} analytic potential at several distances. We
find that Hurly and Mehl's \cite{Hurly} analytic potential should
lead to two $\Phi$ slightly different potentials, as they are shown
in TABLE II. These two potentials are significantly different from
the author's results, but, they are very good consistent with the
most recently SAPT based analytic pair potential\cite{sapt2} for the
helium dimer and other theoretical results\cite{prl,ECG}. So, the
results computed from Hurly and Mehl's \cite{Hurly} analytic
potential in their paper are inaccurate. And near all of their
FN-DMC results (except R=5.6 a.u) are very different from these
analytic potentials and well published theoretical results. All the
aboving results are listed in TABLE II.

In summary, we conclude with pointing out main error in Springall
and co-workers's original paper. They incorrectly computed analytic
potential for the helium dimer from Hurly and Mehl\cite{Hurly}. The
fixed node Diffusion Quantum Monte Carlo calculations in their paper
should be wrong. Thus, their results are in very error, it was
proposed to avoid, their Quantum Monte Carlo simulations should be
revised, a more careful and responsible study is needed.

\textbf{Acknowledgement: }We thank Prof. Krzysztof Szalewicz for
their fortran program for calculating the fit helium dimer
potential.

\begin{table}
\begin{ruledtabular}
\begin{center}
\caption{Comparison of selected predictions of the helium-helium
interaction at R=4.5 bohr. Energies in Kelvin.}\label{table.1}
\renewcommand{\thefootnote}{\thempfootnote}
\begin{tabular}{cc}
 Method & Interaction energy
\\ \hline Fixed node DMC\footnote{The Springall and co-workers's FN-DMC result}      & 49.76$\pm$0.29
\\        Green function QMC\cite{GQMC}      & 60.0
\\ Analytical potential\cite{prl}            & 60.44
\\ r12-MR-ACPF \cite{ACPF}                   & 58.49
\\ CCSD(T)+FCI correction\cite{CCFCI}        & 59.54
\\  SAPT \cite{SAPT}                         & 58.037
\\ CCSD(T)\cite{sapt2}                       & 59.470
\\ SAPT \cite{sapt2}                         & 58.371
\\ CCSDT(Q)/CBS\cite{CCSDTQ}                 & 58.397
\\ CCSD(T)-R12 \cite{ccr12}                  & 59.543
\\ ECG \cite{ECG}                            & 58.517
\\ CCSD(T)/CBS+FCI \cite{ccsdt}              & 58.407
\\ EQMC \cite{EQMC1}                         & 58.3
\end{tabular}
\end{center}
\end{ruledtabular}
\end{table}

\begin{table}
\caption{Interaction energy values of the helium dimer at different
values of the atom separation. The interaction energies are given in
K.}
\begin{ruledtabular}
\begin{tabular}{lccccccl}
\multicolumn{1}{c}{R(bohr)} &
\multicolumn{1}{c}{FN-DMC\footnotemark[1]} &
\multicolumn{1}{c}{Hurly \footnotemark[2] }&
\multicolumn{1}{c}{Hurly \footnotemark[3]} &
\multicolumn{1}{c}{Hurly \footnotemark[4]} & \multicolumn{1}{c}{SAPT
FIT\footnotemark[5]} & \multicolumn{1}{c}{FN-RMC\footnotemark[6]} &
\multicolumn{1}{c}{Other results}
\\
\hline
 0.9 & 314409.15$\pm$3.79 & 319211.08 & 348590.96  & 348588.32  & 350152.17  & 350482.26 & 350489.607\footnotemark[7]\\
 1.9 &  45959.52$\pm$4.11 &  45840.26 &  44863.43  &  44862.40  &  44847.24  &  44852.74 &\\
 2.3 &  20159.33$\pm$0.64 &  20108.76 &  18728.15  &  18727.51  &  18716.36  &  18718.39 &\\
 2.8 &   5575.28$\pm$5.37 &   5549.73 &    6021.75 &   6021.39  &   6020.67  &   6021.23 &\\
 3.2 &   2271.94$\pm$1.45 &   2259.18 &    2333.37 &   2333.15  &   2333.58  &   2334.11 &\\
 3.8 &    537.78$\pm$1.77 &    535.64 &     508.66 &    508.55  &    508.42  &    508.75 &\\
 4.3 &    102.24$\pm$0.32 &    100.67 &     118.13 &    118.07  &    117.93  &    118.01 &\\
 4.5 &     49.76$\pm$0.29 &    50.538 &     58.547 &    58.502  &     58.406 &     58.47 &\\
 4.7 &     19.49$\pm$0.27 &    21.020 &     24.312 &    24.276  &     24.220 &     24.229 &\\
 5.6 &    -10.89$\pm$0.17 &   -11.05  &    -10.9957&   -11.0092 &    -11.0048&    -11.003 &\\
 5.9 &    -10.31$\pm$0.15 &   -10.40  &    -10.1793&   -10.1893 &    -10.1865&    -10.184 &\\
 7.4 &     -2.57$\pm$0.18 &    -3.412 &     -3.3297&    -3.3322 &     -3.3330&     -3.332 &\\
\footnotetext[1]{Springall and co-workers's FN-DMC results
\cite{Springall}} \footnotetext[2]{Springall and co-workers'
inaccurately computed results from Hurly et al's analytic potential
\cite{Hurly}} \footnotetext[3]{our re-computed results, attractive
interaction coefficients for helium atoms with $^{4}He$ nucleii
\cite{Hurly}} \footnotetext[4]{our re-computed results, attractive
interaction coefficients for helium atoms with $^{\infty}He$ nucleii
\cite{Hurly}} \footnotetext[5]{Ref.\cite{sapt2}}
\footnotetext[6]{Our results, to be submitted . (see:
http://arxiv.org/abs/1001.3268)} \footnotetext[7]{Ref.\cite{ECG}}
\end{tabular}
\label{}
\end{ruledtabular}
\end{table}

\end{document}